\documentclass{PoS}

\pdfoutput=1

\usepackage{amsmath}
\usepackage{amstext}
\usepackage{amsfonts}
\usepackage{amssymb}

\title{
\vspace*{-2cm}{\large \normalfont \hfill DESY 11-237, JLAB-THY-12-1410, SFB/CPP-11-75}\\\vspace*{1cm}
Excited State Effects in Nucleon Matrix Element Calculations
\begin{center}
 \includegraphics[scale=0.2]{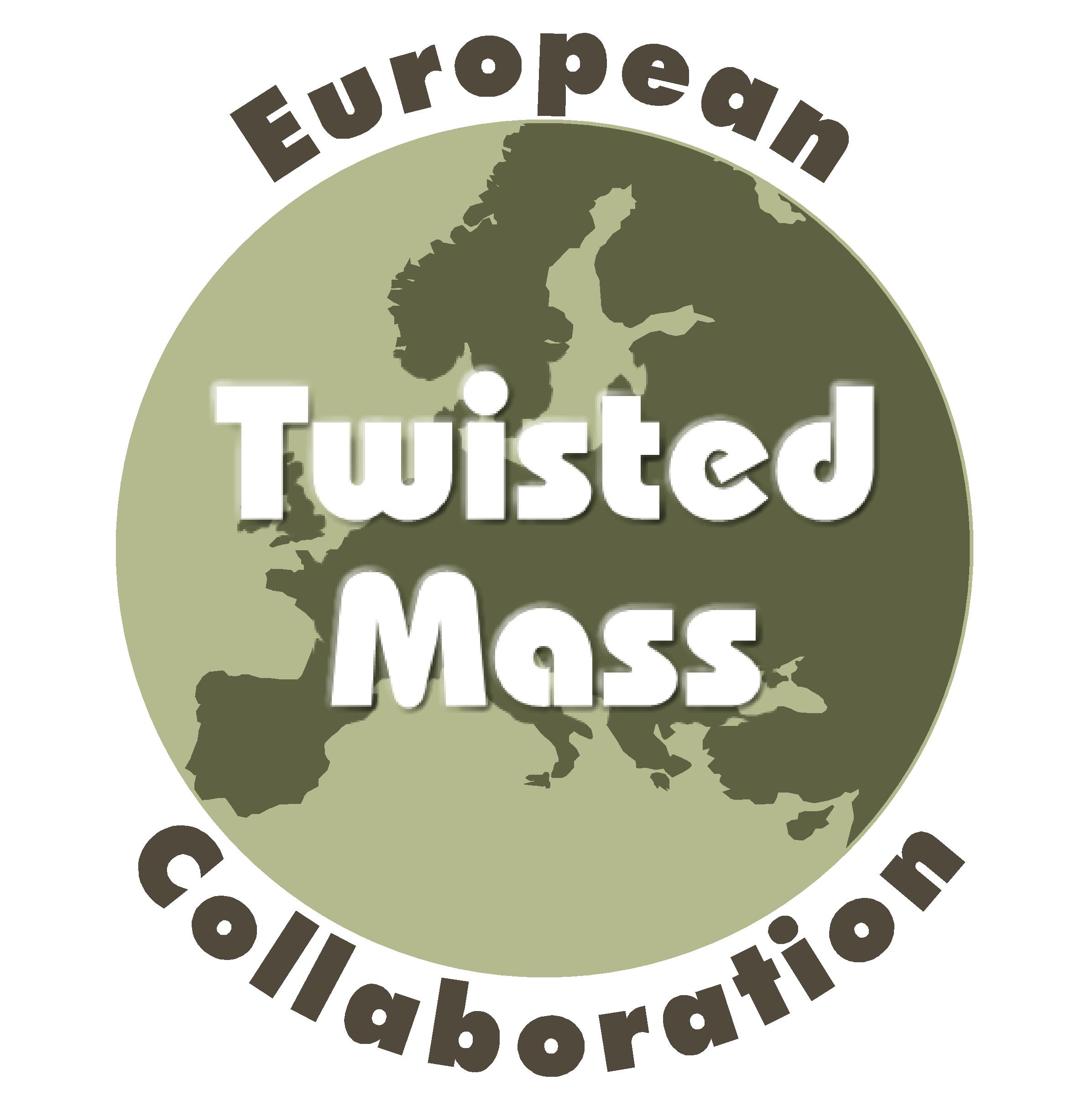}
\end{center}
}

\ShortTitle{Excited State Effects in Nucleon Matrix Element Calculations}

\author{
Constantia Alexandrou$\,^{b,c}$,
Martha Constantinou$\,^b$,
\speaker{Simon Dinter}$\,^a$,
Vincent Drach$\,^a$,
Karl Jansen$\,^a$,
Theodoros Leontiou$\,^d$ and
Dru B.\ Renner$\,^e$\\
\llap{$^a$}{NIC, DESY Zeuthen, Platanenallee 6, D-15738 Zeuthen, Germany\\}
\llap{$^b$}{Department of Physics, University of Cyprus, P.O. Box 20537, 1678 Nicosia, Cyprus\\}
\llap{$^c$}{Computation-based Science and Technology Research Center,
The Cyprus Institute, 15 Kypranoros Str., 1645 Nicosia, Cyprus\\}
\llap{$^d$}{General Department, Frederick University, 1678 Nicosia, Cyprus\\}
\llap{$^e$}{Jefferson Lab, 12000 Jefferson Avenue, Newport News, VA 23606, USA\\}
E-mail: \email{simon.dinter@desy.de}
}

\abstract{We perform a high-statistics precision calculation of nucleon matrix elements 
using an open sink method allowing us to explore a wide range of sink-source time separations. 
In this way 
the influence of excited states of nucleon matrix elements can be studied. 
As particular examples we present results for the nucleon axial charge $g_A$ 
and for the first moment of the isovector unpolarized parton distribution 
$\langle x\rangle_{u-d}$. 
In addition, we report on preliminary results using the generalized 
eigenvalue method for nucleon matrix elements. 
All calculations are performed using $N_f=2+1+1$ maximally twisted mass Wilson fermions.}

\FullConference{XXIX International Symposium on Lattice Field Theory \\
 July 10 -- 16 2011\\
 Squaw Valley, Lake Tahoe, California}

\graphicspath{{graphics/}}

\begin{document}
\section{\label{sec:int}Introduction}
\begin{figure}[htb]
\centering
\includegraphics[width=0.6\textwidth]{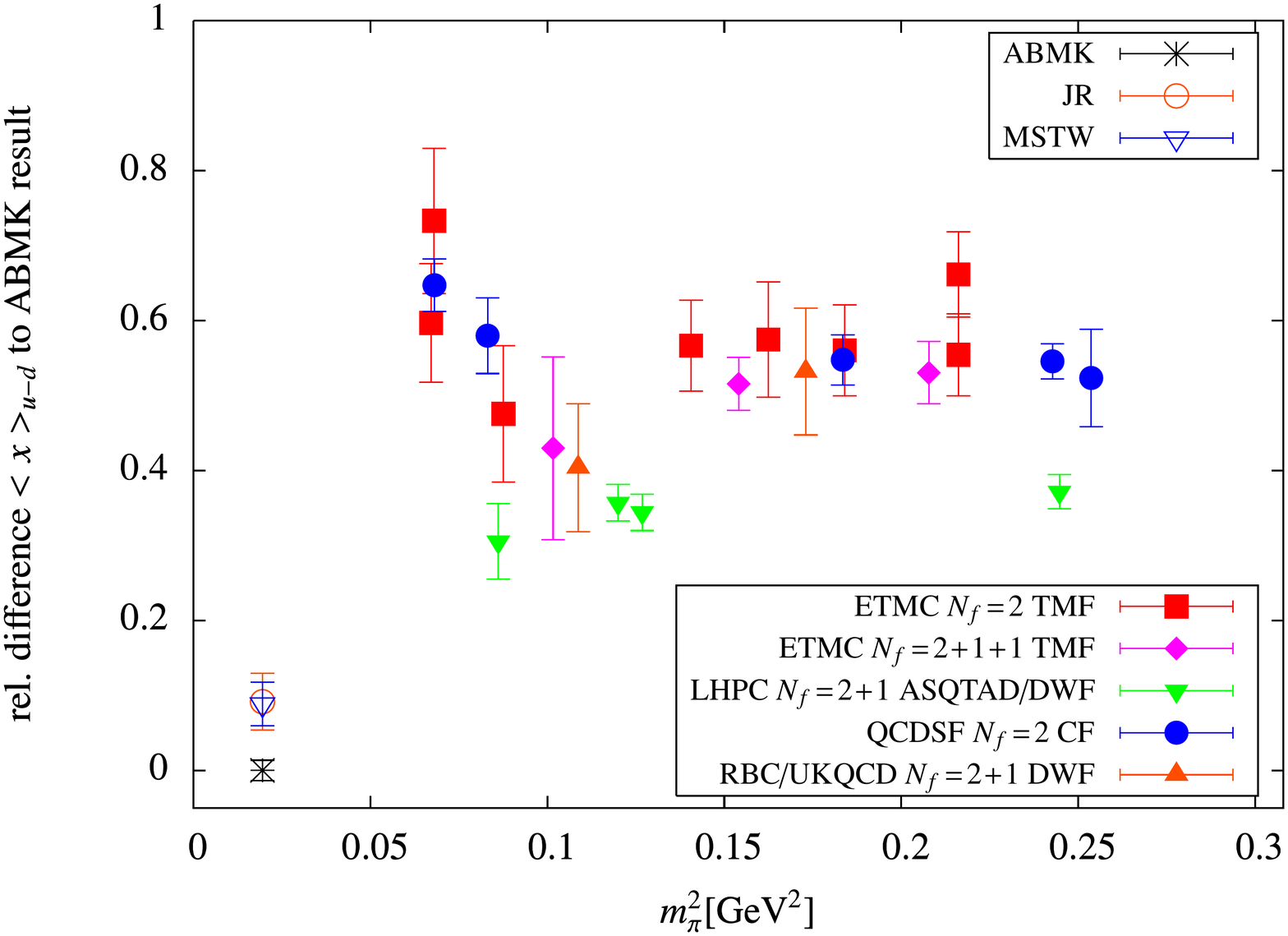}
\vspace{-4mm}
\caption{\label{fig:results_averageX}
With the filled symbols we show the deviation of the values of $<x>_{u-d}$ 
obtained from lattice calculations 
using twisted mass fermions (TMF)~\cite{Alexandrou:2011nr,Dinter:2011jt} (ETMC), 
a hybrid action of domain wall fermions (DWF) on a
stagerred sea~\cite{Bratt:2010jn} (LHPC),
Clover fermions~\cite{Pleiter:2011gw,Sternbeck:2011pc} (QCDSF) and DWF~\cite{Aoki:2010xg} (RBC/UKQCD). 
The open symbols show the deviation of the phenomenological 
results~\cite{JimenezDelgado:2008hf} (JR) and~\cite{Martin:2009bu} (MSTW) 
as compared to ABMK~\cite{Alekhin:2009ni}.}
\vspace{-2mm}
\end{figure}

\vspace{-3mm}
Quantities related to nucleon structure  are  
among the most challenging  that can be computed within 
lattice QCD.
During the past few years,
there has been significant progress in these calculations: 
quark masses used nowadays are close to the physical ones, 
lattice spacings are small enough to allow for a controlled extrapolation 
to the continuum limit 
and volumes are sufficiently large to suppress finite volume effects. 
But still even for some simple observables like the nucleon axial charge
$g_A$ or the average momentum of the unpolarized isovector parton distribution
$\langle x\rangle_{u-d}$ the values obtained from lattice calculations differ from their corresponding experimental values.
In Fig.~\ref{fig:results_averageX} we show the relative deviation 
of lattice results for $<x>_{u-d}$ from the value obtained in Ref.~\cite{Alekhin:2009ni} (ABMK). 
We also show the  deviation of results from different
phenomenological extractions from deep inelastic scattering experiments
as compared to that of ABMK.
The deviations of the phenomenologically determined 
values for $<x>_{u-d}$ between the different groups 
might be interpreted as a systematic error related to different data sets and
different analysis/fitting strategies. However, as can be seen,
the spread in the phenomenological number is much less than the 
relative deviation between lattice data and phenomenology. 
One possible reason for the discrepancy can be 
systematic effects that enter in the lattice calculations.
Therefore the latter have to be thoroughly studied, which is a current
important activity of many lattice collaborations.
In these proceedings we focus on one potential systematic effect,
namely excited state contributions.
Furthermore we restrict ourselves to the two  observables mentioned before,
$g_A$ and  $\left\langle x \right\rangle_{u-d}$, that are extracted from
the matrix element at zero momentum transfer.
These quantities are calculated in lattice QCD by studying the 
asymptotic behavior of suitable ratios
of 3-point and 2-point correlation functions for large Euclidean time separations.

Several studies of $g_A$ and  $\left\langle x \right\rangle_{u-d}$   
at different values of the lattice spacing and 
various volumes~\cite{Alexandrou:2011nr,Aoki:2010xg,Bratt:2010jn,Pleiter:2011gw} 
indicate that for the pion masses considered, 
discretization and finite volume effects are
likely not sufficient to explain the disagreement of
the lattice calculations with the current experimental value of $g_A$ 
and the value of $\left\langle x \right\rangle_{u-d}$ from global analyses.
Moreover, recent preliminary results~\cite{Pleiter:2011gw} shows that the discrepancy 
persists even for pion masses almost as small as the physical one. 

In order to clarify whether and to what extent
excited state contributions affect the lattice calculation 
of matrix elements, we have performed a dedicated high-precision calculation  
of the 3-point functions for $g_A$ 
and $\left\langle x \right\rangle_{u-d}$. To be more precise, we have used 
roughly $7500$ measurements for $g_A$ and about
$23000$ for $\left\langle x \right\rangle_{u-d}$, 
which enabled us to calculate the correlation functions 
needed for $g_A$ and  $\left\langle x \right\rangle_{u-d}$
to sufficient accuracy even for large Euclidean time separations.
This allowed us to study possible excited state contributions.
%
%
\section{\label{sec:techniques}Lattice techniques and details}
\vspace{-3mm}
In this work, we have employed twisted mass Wilson fermions~\cite{Frezzotti:2003ni} 
at maximal twist. This lattice discretization has the advantage 
of an {\em automatic} $O(a)$-improvement, thus removing the necessity of additional
operator specific improvements. 
We use gauge field configurations generated by   
the European Twisted Mass Collaboration (ETMC)
with $N_f=2+1+1$ flavors and refer to Ref.~\cite{Baron:2010bv} for further details 
of our lattice formulation. 

To understand how the excited state contributions emerge from the 
calculation of nucleon matrix elements
the main observation needed here is that to leading order we find 
two additional time dependent contributions
for a matrix element $\left\langle N \right| \mathcal{O} \left| N \right\rangle$, namely
\begin{align}
 \quad \left\langle N \right| \mathcal{O} \left| N \right\rangle_{\text{lattice}}
   &=
 \quad \left\langle N \right| \mathcal{O} \left| N \right\rangle 
    +  A \: \exp\left(-\Delta M t^\prime\right)
    +  B \: \exp\left[-\Delta M (t - t^\prime)\right].
\end{align}
Here $\Delta M$ is the mass gap between the nucleon ground state and the 
first excited state and $A$ and $B$ are constants depending on the
particular choice of the lattice nucleon creation and annihilation operator,
respectively.
For details  we refer the reader
to Ref.~\cite{Dinter:2011sg}.
For our precision calculation we restrict ourselves to only one ensemble
with a pion mass of $m_\pi\approx 380~\mathrm{MeV}$ and a
lattice spacing of $a \approx 0.078~\text{fm}$.
This pion mass is chosen sufficiently small in order to be relatively 
close to the physical pion mass on the one hand but still large enough to
ensure that finite size effects can be safely neglected. 
Also, for such a pion mass, the calculations of the 
propagators do not require large computer resources,
thus enabling us to collect the statistics 
necessary for the high precision we are aiming at.
We expect discretization effects to be small due to the automatic $O(a)$-improvement,
which is confirmed by previous calculations of nucleon matrix elements 
at three different lattice spacings smaller than $0.1~\mathrm{fm}$ 
\cite{Alexandrou:2011nr,Alexandrou:2010hf,Alexandrou:2010cm} 
using $N_f=2$ maximally-twisted-mass fermions.
Let us point out in particular, that even though 
this analysis was performed using twisted-mass fermions, 
the important properties of excited state contributions
are expected to be universal and independent of the particular
lattice discretization used. 
Therefore we expect our conclusions to hold qualitatively also 
for different lattice actions.

For the precision calculation of the 3-point functions 
performed in this work we employed a method that is different from 
the standard sequential method, where the sink time slice is fixed.
We fixed the time slice of the operator insertion as well as the operator to 
obtain a result for all sink time slices. 
This enables us to study the source-sink separation dependence of a particular matrix element.
Note that in general the standard calculation is more desirable since 
it does not depend on the operator itself, 
whereas in the open sink method we have to restrict to a single operator.
The interested reader is referred to 
Ref.~\cite{Dinter:2011sg} for a more detailed discussion.

As stated in the introduction, 
we concentrate on two relatively simple but nonetheless 
phenomenologically relevant quantities.
The first is the nucleon axial charge, $g_A$, playing   
an important role in the beta decay of the 
neutron and appearing as a low energy constant in effective chiral Lagrangians.
It has been experimentally measured precisely and 
it is also straightforward to calculate in lattice QCD.
However, the values obtained from various lattice calculations 
are typically $5\%$ to $10\%$ below the experimental result \cite{Nakamura:2010zz} while having themselves a 
statistical accuracy of the order of $1\%$.
The second observable is the lowest non-trivial moment of the unpolarized 
parton distribution function in isovector flavor combination,
$\left\langle x \right\rangle_{u-d}$. 
It is determined phenomenologically from 
a global analysis of deep inelastic scattering data,
and the discrepancy between the phenomenological 
and lattice values is even larger, about $50\%$ to $60\%$, 
as can be seen in Fig.~{\ref{fig:results_averageX}}.
We would also like to stress that non-perturbative renormalization is employed
for the bare matrix elements. The corresponding renormalization factors are calculated in the 
$\text{RI}^\prime\text{MOM}$ scheme and are matched to the $\overline{\text{MS}}$ 
scheme at a scale of $(2~\text{GeV})^2$. For more details we refer to Refs.~\cite{Alexandrou:2010me,Dinter:2011jt}. 
The values of the renormalization constants relevant here are $Z_A = 0.774$ for the 
renormalization of the bare $g_A$ and $Z_{\langle x\rangle}= 0.998$ for the renormalization 
of $\left\langle x \right\rangle_{u-d}$~\cite{Alexandrou:2010kv}. 
%
%
\section{\label{sec:results}Results}
\begin{figure}[ht]
\includegraphics[width=0.49\textwidth]{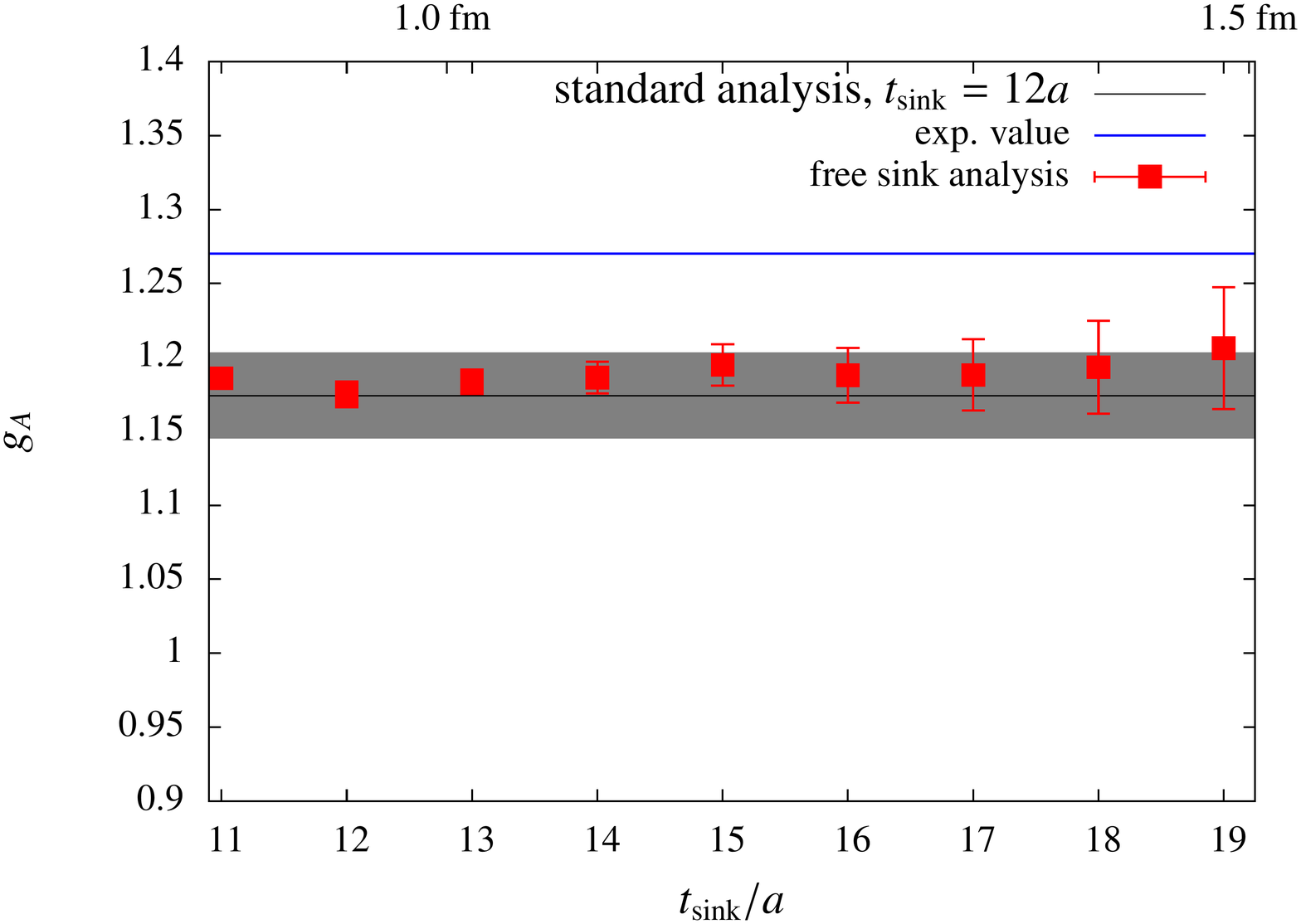}
 \includegraphics[width=0.49\textwidth]{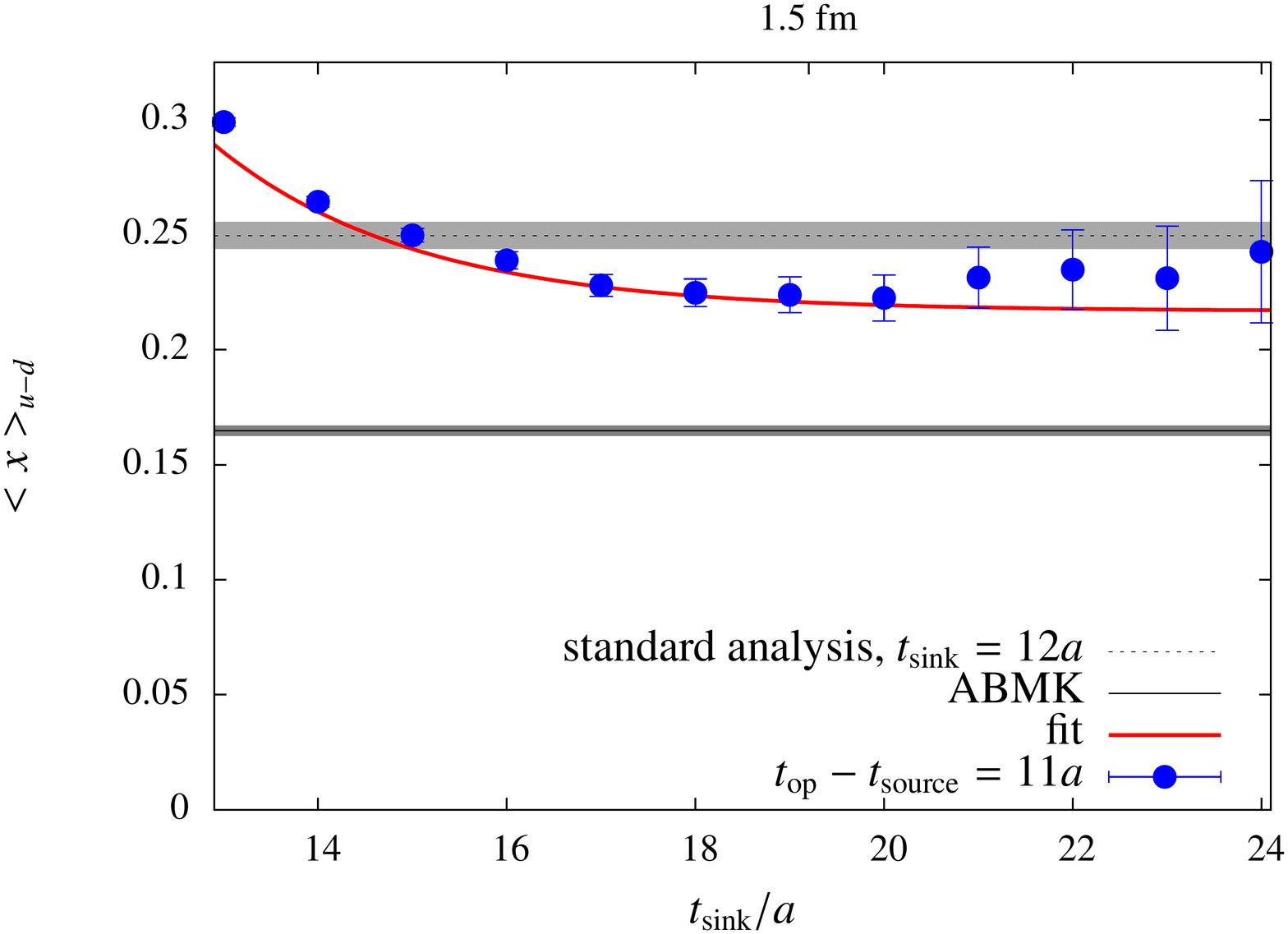}
\vspace{-3mm}
 \caption{\label{fig:gA_and_averageX_open_sink}
Left panel: Results for $g_A$ for a range of source-sink separations 
obtained from the open sink analysis on one $N_f=2+1+1$ ensemble.
The light gray band indicates the result obtained from the fixed sink method  
using a source-sink separation of $12a$ and the dark gray band shows the experimental value.
Right panel: 
$\left\langle x \right\rangle_{u-d}$ for a range of 
source-sink separations obtained by means of the open sink method.
The value (including errors) obtained from the fixed sink method
using a source-sink separation of $12a$ is indicated by the light gray band.
The phenomenologically extracted value is shown with the dark gray band. 
The blue solid line corresponds to a fit described in the text.}
\vspace{-1mm}
\end{figure}

\begin{figure}[htb]
\vspace{-2.5cm}
\centering
\includegraphics[width=0.8\textwidth]{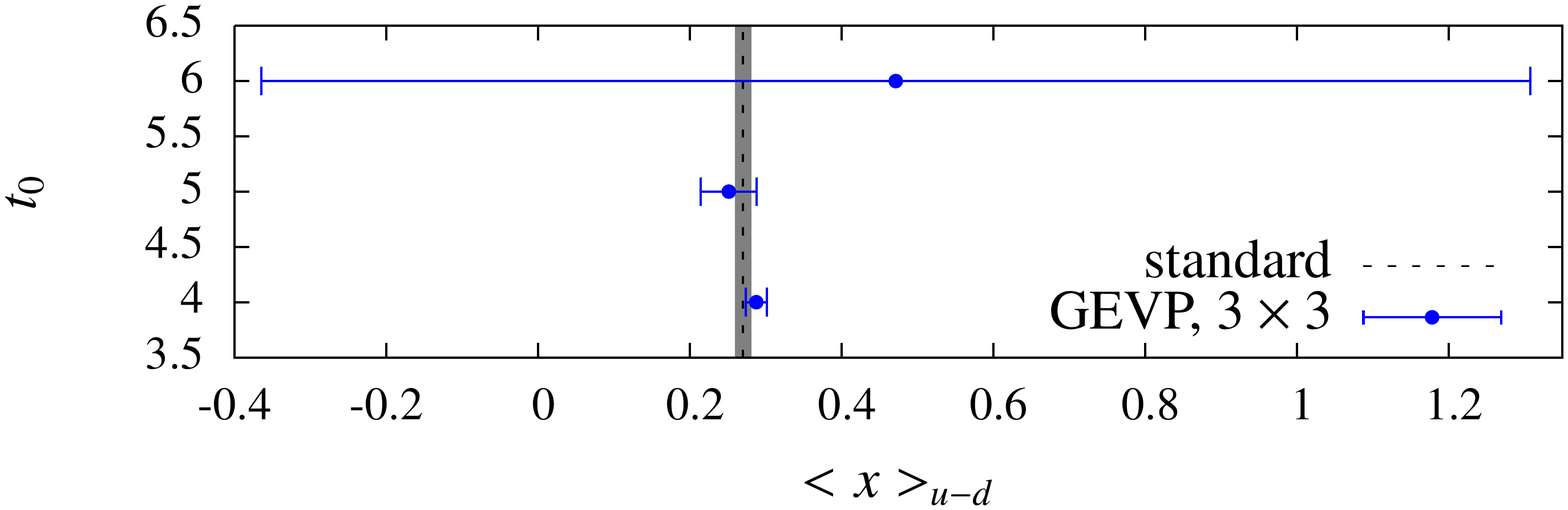}
\vspace{-2.6cm}
\caption{\label{fig:plateau_and_GEV}
Comparison between the value obtained from a standard calculation 
of $\left\langle x\right\rangle_\text{u-d}$ 
using a fixed source-sink separation of $12a$ and preliminary results from a GEVP analysis 
using the same source-sink separation for a range of $t_0$.}
\vspace{-5mm}
\end{figure}

\vspace{-3mm}
We performed an analysis of $g_A$ on a single $N_f=2+1+1$ ensemble 
characterized in Sec.~\ref{sec:techniques} 
using the open sink method. 
The time slice of the operator insertion was fixed to $t'=9a$, chosen
to safely suppress excited state contributions from the source,
as can be verified from the 2-point function.
We have used Gaussian smearing of the quark fields, including APE-smeared gauge links,
in order to improve the overlap with the nucleon ground state.
The result of the analysis using the open sink method is shown in 
Fig.~{\ref{fig:gA_and_averageX_open_sink}} (left). 
In this figure we also compare the values obtained from the open sink analysis 
to the value of a standard analysis with a fixed source-sink separation of 
$12a$ on the same ensemble.
The value of $g_A$ does not show any dependence on 
the source-sink separation $t$ within statistical accuracy and
thus demonstrates the absence of significant contribution from excited states.
It is worth mentioning that, in order to reach a comparable 
statistical accuracy as the one obtained when using the fixed sink method 
with $t=12a$ with 500 measurements, we had to perform
roughly $7500$ measurements when we take e.g.\ $t=18a$.
We have run the same analysis for $\left\langle x \right\rangle_{u-d}$.
For the open sink method, we have chosen the operator insertion time to be $t^\prime=11a$. 
We expect that for this choice excited state effects 
from the source are sufficiently suppressed.
With a statistics of $23,000$ measurements for $\left\langle x \right\rangle_{u-d}$ 
at a source-sink separation of $t=18a$, 
we could equal the precision of the fixed sink method that was done 
with a source-sink separation of $t=12a$ using $1300$ measurements.
In the right panel of Fig.~\ref{fig:gA_and_averageX_open_sink} 
we plot $\left\langle x \right\rangle_{u-d}$ 
as a function of the source-sink separation $t$.
We also indicate the value obtained from the fixed sink method 
analysis as well as the experimental result from a 
global analysis~\cite{Alekhin:2009ni}.
We observe that the values for $\left\langle x \right\rangle_{u-d}$ 
reach a plateau at larger values of the source-sink separation than what
we have used in the fixed sink method and that the plateau value is shifted. 
Still, despite the fact that the results for larger values of $t$ decrease
they clearly do not reach the phenomenological value. 
In order to estimate the residual dependence on $t$,
we determined the value of $\left\langle x \right\rangle_{u-d}$
for an infinite source-sink separation 
by fitting the expected exponential behavior,
\begin{align*}
\left\langle x \right\rangle_{u-d} + A \exp \left[ -\Delta M\left(t - t'\right) \right] \;,
\end{align*}
to the lattice results with a fixed $t'=11a$. 
The result of this fit is $\left\langle x \right\rangle_{u-d} = 0.22(1)$. This value 
is $12\%$ lower than the result of $\left\langle x \right\rangle_{u-d} = 0.250(6)$, 
obtained using $t=12a$ in the fixed sink method. 
We estimated the error of the fit by varying the fit range and 
by comparing the results obtained by using a fixed parameter $\Delta M$ 
as well as by including $\Delta M$ in the fit.
%
%
\begin{figure}[h]
\begin{minipage}{0.49\linewidth}
\includegraphics[width=\linewidth]{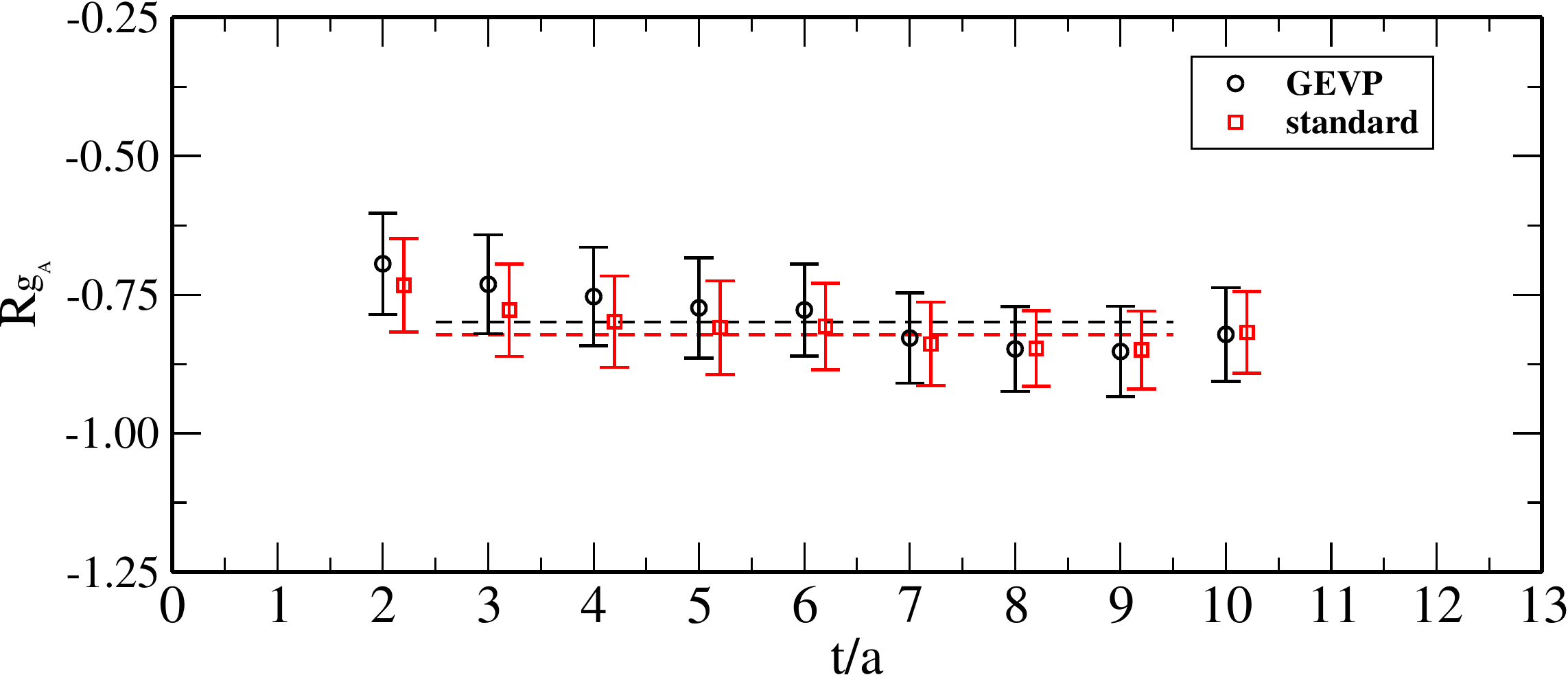}
\end{minipage}
\begin{minipage}{0.49\linewidth}
\includegraphics[width=\linewidth]{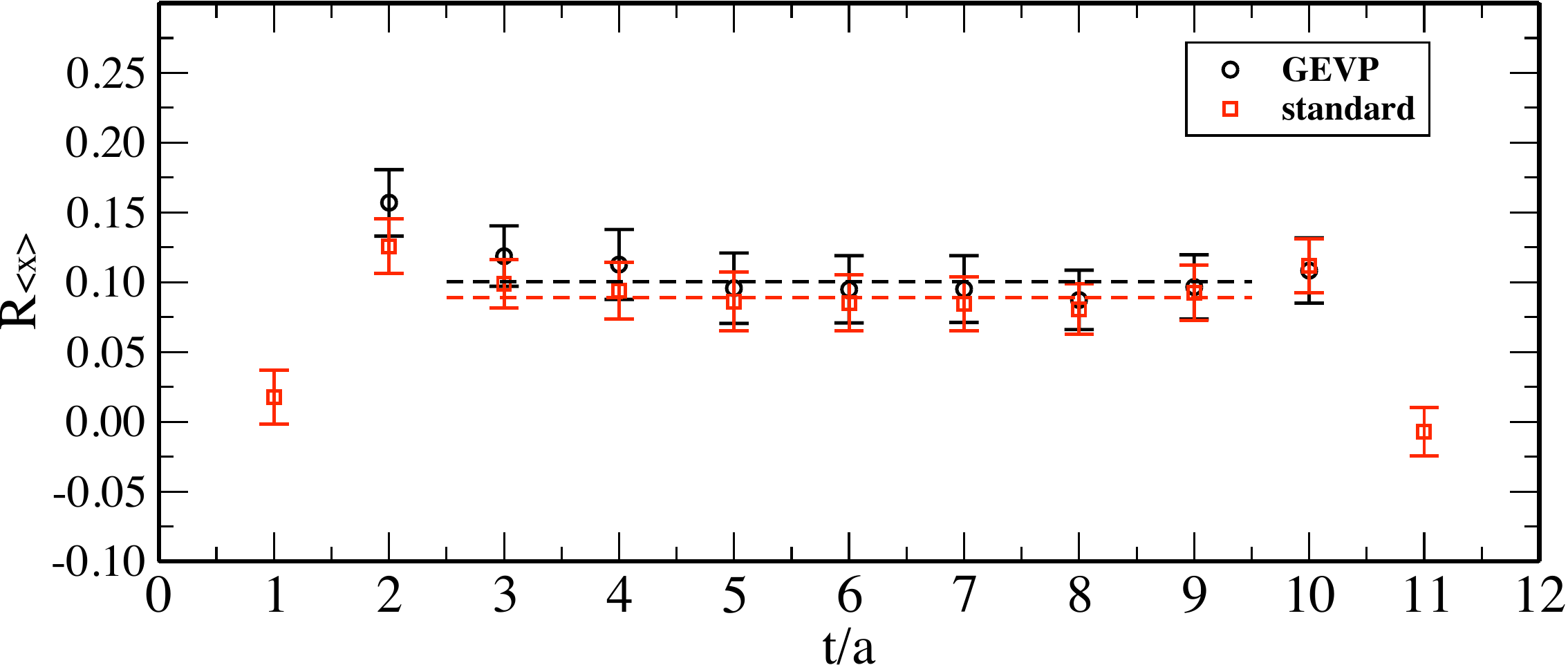}
\end{minipage}
\caption{\label{fig:GEV} Ratio of 3- to 2-point functions for $g_A$ (left) 
and $\langle x \rangle_{u-d}$ (right) versus the time separation of the operator insertion 
from the source. The sink-source time separation
$t_{\rm sink}=12a$. $N_f=2$ gauge configurations are used with pion mass $\sim 300$~MeV and $a=0.089$~fm.}
\vspace{-1mm}
\end{figure}
We have also used a generalized eigenvalue (GEV) approach 
as first
suggested in \cite{Luscher:1990ck} and further developed and refined
in Refs.~\cite{Blossier:2009kd, Blossier:2010mk,Bulava:2011yz}.
We considered two variational basis:
in the first, two interpolating fields are considered, namely the standard one 
$J_N(x)=\epsilon^{abc}(d^{T\,a}(x)C\gamma_5u^b(x))u^c(x)$ and   
$J^\prime_N(x)=\epsilon^{abc}(d^{T\,a}(x)Cu^b(x))\gamma_5u^c(x)$. 
The latter is known to have  small overlap with the nucleon state but
a large one with the Roper. In Fig.~\ref{fig:GEV} we compare the ratio
of the three-point to the two-point function arising when using the GEV approach
to the one using just the standard $J_N(x)$ interpolating field
 as we vary the time separation of the operator insertion form the source. 
As can be seen, for both  $g_A$ and $\langle x \rangle_{u-d}$ the 
ratios are consistent and lead to the same value for these observables.  
In the second,  three different levels of smearing
are employed to
 calculated a $3\times 3$ correlation matrix.
In Fig.~\ref{fig:plateau_and_GEV} we show the effective plateau matrix elements
for fixed $t_\text{op}=t_\text{sink}/2$ and various $t_0$.
From the preliminary results as shown in Fig.~\ref{fig:plateau_and_GEV}, 
we conclude that within the statistical errors we can achieve 
with our nucleon 3-point functions, the values obtained from the GEV
are fully consistent with our standard method. 
%
%
\section{\label{sec:summary}Summary and Conclusions}
\vspace{-3mm}
We have performed precision calculations of 
$g_A$ and $\left\langle x \right\rangle_{u-d}$ for a single ensemble 
of gauge field configurations corresponding to 
a pion mass of about $380~\text{MeV}$ with $N_f=2+1+1$ dynamical fermions 
employing a non-perturbative renormalization. 
We have studied the dependence of these quantities 
on the source-sink separation in order to 
assess the influence of excited states on the 
current lattice results for
$g_A$ and $\left\langle x \right\rangle_{u-d}$. 
This is particularly important given that excited states may play a role in
explaining the presently observed discrepancy between lattice computations and 
phenomenological evaluations of several important observables 
related to nucleon structure.

We find that for the here considered pion mass of about
$380~\text{MeV}$ and lattice spacing of $a\approx0.078~\mathrm{fm}$, 
the contamination of excited states 
is negligible for $g_A$, but for 
$\left\langle x \right\rangle_{u-d}$ we observe
an effect that is of the order of $10\%$ compared to 
our previous calculations,
in which the source-sink separation has been fixed to 
approximately $1$~fm. 
This is an effect larger than 
the finite volume and lattice spacing effects found 
at this value of the pion mass, volume and lattice spacing.
Moreover, this study shows that excited state contributions are
operator dependent and should be investigated separately for each operator.

Motivated by a study where variational methods have been 
employed in the calculation of the $B^\ast B\pi$ coupling \cite{Bulava:2011yz}, 
we applied the generalized eigenvalue method \cite{Luscher:1990ck,Blossier:2009kd} to calculate these quantities. 
The results of this analysis were consistent with the standard method. 
This indicates that in order to extract excited state effects, for which the GEVP method is expected
to be very efficient, a significantly higher accuracy of the lattice data are needed. 
%
%
\section*{Acknowledgments}
\vspace{-3mm}
This work is coauthored in part by Jefferson Science Associates, 
LLC under U.S. DOE Contract No. DE-AC05-06OR23177.
It was partly supported by the Cyprus Research Promotion Foundation 
under contracts TECHNOLOGY/$\Theta$E$\Pi$I$\Sigma$/0308(BE)/17,
and $\Delta$IAKPATIKE$\Sigma$/KY-$\Gamma$A/0310/02 and
by the Research Executive Agency of the European Union 
under Grant Agreement number PITN-GA-2009-238353 (ITN STRONGnet).
It is additionally supported in part by the DFG Sonderforschungsbereich Transregio SFB/TR9.
Numerical calculations have been performed using HPC resources 
provided by J\"ulich Supercomputing Centre at the research center in J\"ulich.
\bibliography{proceedings}
\end{document}